\documentclass[a4paper,11pt]{article}
\pdfoutput=1 
\usepackage{jheppub} 
\newcommand{\mathsym}[1]{{}}

\newcommand{\ba}{\begin{array}}
\newcommand{\ea}{\end{array}}
\newcommand{\be}{\begin{equation}}
\newcommand{\ee}{\end{equation}}
\newcommand{\beqa}{\begin{eqnarray}}
\newcommand{\eeqa}{\end{eqnarray}}
\newcommand{\degree}{\ensuremath{^\circ}}
\def\mt{$\mu$-$\tau$ }
\def\mnuf{${\cal M}_{\nu f}$}

\title{\boldmath How good is $\mu$-$\tau$ symmetry after results on non-zero $\theta_{13}$ ?}
\author[a]{Shivani Gupta,}
\author[b]{Anjan S. Joshipura}
\author[c]{and Ketan M. Patel}

\affiliation[a]{Department of Physics and IPAP, Yonsei University, Seoul 120-479, Korea.}
\affiliation[b]{Physical Research Laboratory, Navarangpura, Ahmedabad 380009, India.}
\affiliation[c]{Tata Institute of Fundamental Research, Homi Bhabha Road, Colaba, Mumbai 400005, India.}

\emailAdd{shivani@cskim.yonsei.ac.kr}
\emailAdd{anjan@prl.res.in}
\emailAdd{ketan@theory.tifr.res.in}

\preprint{TIFR/TH/13-03}

\abstract{Viability of the $\mu$-$\tau$ interchange symmetry imposed as an approximate symmetry (1) on the neutrino
mass matrix ${\cal M}_{\nu f}$  in the flavour basis (2) simultaneously on the charged lepton mass matrix $M_l$ and the
neutrino mass matrix $M_\nu$ and (3) on the underlying Lagrangian is discussed in the light of recent observation of a
non-zero reactor mixing angle $\theta_{13}$. In case (1), $\mu$-$\tau$ symmetry breaking may be regarded as small (less
than 20-30\%) only for the inverted or quasidegenerate neutrino mass spectrum and the normal hierarchy would violate it
by a large amount. The case (2) is more restrictive and the requirement of relatively small breaking
allows only the quasidegenerate spectrum. If neutrinos obtain their masses from the type-I seesaw
mechanism then small breaking of the $\mu$-$\tau$ symmetry in the underlying Lagrangian may result
in a large breaking in ${\cal M}_{\nu f}$  and even the hierarchical neutrino spectrum may also be
consistent with mildly broken $\mu$-$\tau$ symmetry of the Lagrangian. Neutrinoless double beta
decay provides a good means of distinguishing above scenarios. In particular, non-observation of
signal in future experiments such as GERDA would rule out scenarios (1) and (2).}

\begin{document}
\maketitle
\flushbottom

\section{Introduction}
After a conclusive evidence of a non-zero $\theta_{13}$ by  several reactor neutrino experiments
\cite{experiments} disfavoring $\theta_{13}=0$ with $\Delta\chi^2 \approx 100$ in a global analysis
\cite{fit:valle-fogli, GonzalezGarcia:2012sz}, it is more meaningful to turn the theoretical search
for a symmetry leading to zero $\theta_{13}$ to a systematic study of effects of perturbations on it
(see the recent reviews \cite{recent_review} and references therein) or to a search for an
alternative symmetry which can predict nonzero $\theta_{13}$. Some of the specific symmetries which
ensure this are identified in the literature \cite{symmetry}. The effect of perturbations to
underlying symmetry giving $\theta_{13}=0$ can be studied more generally \cite{Grimus:2004cc,Mohapatra:2004mf}
purely at the phenomenological level. Irrespective of any underlying model, one can define an effective $Z_2$ symmetry
which is both necessary and sufficient for obtaining $\theta_{13}=0$ \cite{Grimus:2004cc}. This is generated by the
transformation $S$ :

\be
\label{s}
S=\left(
\ba{ccc}
1&0&0\\
0&\cos2\theta_{23}&\sin 2\theta_{23}\\
0&\sin2 \theta_{23}&-\cos 2\theta_{23}\\
\ea \right) ~,\ee
where $\theta_{23}$ denotes the atmospheric mixing angle. Invariance of the
neutrino mass matrix \mnuf~in flavour basis under $S$ leads to vanishing
$\theta_{13}$. A well-motivated special case of $S$ is the celebrated \mt
symmetry \cite{mt} which is obtained from Eq. (\ref{s}) when $\theta_{23}=\pi/4$ :
\be \label{s2}
S_2=\left(
\ba{ccc}
1&0&0\\
0&0&1\\
0&1&0\\ \ea
\right)~. \ee

We will concentrate here on this specific symmetry and consider two different scenarios. In the
first, we assume that \mt is an effective symmetry of \mnuf. This symmetry may be accidental or a
consequence of some other (e.g. $D_4$ \cite{Grimus:2003kq}) broken symmetry. In such a situation,
the (diagonal) charged lepton mass matrix breaks \mt symmetry. In an alternative scenario, we
regard \mt symmetry as more fundamental and impose it as an approximate symmetry of both the charged
lepton and neutrino mass matrix. This can arise from \mt symmetry imposed at the Lagrangian level
itself. Our main aim in this paper is to carry out a detailed quantitative assessment of the
magnitude of the \mt symmetry breaking required in both these scenarios in order to explain the observed value of
$\theta_{13}$. The viability or otherwise of the \mt symmetry is then discussed in the light of such quantitative
analysis. In particular, we observe a close link between the neutrino mass hierarchy and the amount of \mt symmetry
breaking parameters in both these scenarios and find that specific mass hierarchies are preferred by the small
breaking of \mt symmetry in each scenario.

\section{Approximately \mt symmetric \mnuf}
To be specific, we  define \mt symmetry by requiring that the eigenvector of the neutrino mass
matrix \mnuf~in flavour basis corresponding to the heaviest (lightest) mass eigenvalue is given by
\be \label{ev}
\left( \ba{c}
0\\ \pm \frac{1}{\sqrt{2}}\\\frac{1}{\sqrt{2}} \\ \ea \right) \ee   
in case of the normal (inverted) hierarchy in the neutrino masses. This requirement leads to the
following form for \mnuf:
\be \label{mnuf1}
{\cal M}_{\nu f}^0=
\left(
\ba{ccc}
X&A&\mp A\\
A&B & C\\
\mp A &C& B\\
\ea
\right)~.
\ee
These two are special cases of the more general symmetry Eq. (\ref{s}), obtained when
$\theta_{23}=\pm \frac{\pi}{4}$. Since sign of $\theta_{23}$ can be changed  by appropriately
defining CP violating phases and charged lepton mass eigenstates, it is sufficient to consider only 
one of the two and we will choose the  one corresponding to the negative sign in Eq. (\ref{ev}). 
All parameters above are complex but two of them say, $X$ and $C$ can be made real by redefining the
phases of the charged lepton mass eigenstates. ${\cal M}_{\nu f}^0$ is thus characterized by six
real parameters and leads to two predictions among eight\footnote{Dirac phase $\delta$ becomes unphysical
because of the prediction $\theta_{13}=0$.} relevant observables in the neutrino sector.
Note that Eq. (\ref{ev}) is an eigenvector of Eq. (\ref{mnuf1}) with the eigenvalue
$B\pm C$. If this eigenvalue corresponds to the heaviest (lightest) mass eigenstate in case of the
normal (inverted) mass hierarchy then the said two predictions are: $\theta_{13}=0$ and
$\theta_{23}=\pi/4$. If this is not the case then one obtains $\theta_{12}=0$ or $\pi/2$ instead
of $\theta_{13}=0$. This case is also of interest as a small perturbation to it may result in a
large $\theta_{12}$, see \cite{Liao:2012xm} for a discussion of this case.  Here, we
only consider the case which predicts $\theta_{13}=0$ in the exact \mt symmetric limit.

\mt symmetry is also defined in the literature  in a generalized sense which combines
ordinary interchange of $\mu$-$\tau$ symmetry with the CP \cite{gencp}. In this case, Eq.
(\ref{mnuf1}) gets replaced by 
\be \label{mnuf2}
{\cal M}_{\nu f}^0=
\left(
\ba{ccc}
X&A&A^*\\
A&B & C\\
A^* &C& B^*\\
\ea
\right)~.
\ee
Here $X$ and $C$ are forced to be real. It is then possible to remove an additional phase from
either $B$ or $A$ by redefining the charged lepton mass eigenstates without affecting the reality of
$X$ and $C$. Above ${\cal M}_{\nu f}^0$ is thus characterized by five real parameters and leads to
four predictions among the nine observables. These correspond to two trivial Majorana phases and the
relations:
\be
\theta_{23}=\frac{\pi}{4}~~~{\rm and}~~~
{\rm Re}(\cos\theta_{12}\sin\theta_{12}\sin\theta_{13}e^{i\delta})=0~.\ee
Unlike in Eq. (\ref{mnuf1}), the ${\cal M}_{\nu f}^0$ as given in Eq. (\ref{mnuf2})
is phenomenologically allowed and the generalized \mt symmetry can still remain an exact symmetry.
We will concentrate here on the \mt symmetry as in Eq. (\ref{mnuf1}) and discuss effect of
perturbation on it. The \mt symmetry in ${\cal M}_{\nu
f}$ implies equalities: $({\cal M}_{\nu f})_{12}=({\cal M}_{\nu
f})_{13}$ and $({\cal M}_{\nu f})_{22}=({\cal M}_{\nu f})_{33}$. It is
thus natural to characterize its breaking in terms of two complex parameters
defined as follows:
\be \label{defe1e2}
\epsilon_1\equiv \frac{({\cal M}_{\nu f})_{12}-({\cal M}_{\nu f})_{13}}{({\cal
M}_{\nu f})_{12}+({\cal M}_{\nu f})_{13}} ~;~~
\epsilon_2\equiv \frac{({\cal M}_{\nu f})_{22}-({\cal M}_{\nu f})_{33}}{({\cal
M}_{\nu f})_{22}+({\cal M}_{\nu f})_{33}} ~.
\ee
We would define approximate \mt symmetry as the one in which the absolute values of the above
dimensionless parameters $\ll 1$. Let us note that
\begin{itemize}
\item $\epsilon_{1,2}$ characterize the most general breaking of the \mt symmetry and all other
elements of an arbitrary perturbation matrix to Eq. (\ref{mnuf1}) can be absorbed in ${\cal M}_{\nu
f}^0$ given in Eq. (\ref{mnuf1}).
\item  One could have normalized the \mt breaking denominators in the above equation with a
different quantity, {\it e.g.} the largest neutrino mass. Such a definition would be less
conservative and may imply small $\epsilon_{1,2}$ even when percentage deviation in the differences
in the numerator is very large. 
\end{itemize}
One can relate the parameters $\epsilon_{1,2}$ to observable quantities in a straightforward manner.
 Before the measurement of $\theta_{13}$, such an approach was taken in \cite{Abbas:2010jw} to analyze 
the effects of deviations of the lepton mixing from its tri-bimaximal values on the structure of neutrino mass 
matrix. This includes \mt symmetry as a special case. We concentrate here only on \mt symmetry of ${\cal
M}_{\nu f}$ and discuss its viability in three physically distinct situations using the precise measurements of
$\theta_{13}$.
The most general neutrino mass matrix \mnuf~can be written after appropriate rephasing of
charged lepton mass eigenstates as 
\be
\label{mnufgeneral}
{\cal M}_{\nu f}=U^*{\rm Diag.}(m_1,m_2,m_3) U^\dagger~,\ee
where 
\be
\label{upmns}
U=
\left(
\ba{ccc}
c_{13}c_{12}&-c_{13}s_{12}&- s_{13}e^{-i \delta}\\
c_{23} s_{12} - c_{12}  s_{13} s_{23} e^{i\delta}&c_{12} c_{23} +  s_{12} s_{13}
s_{23} e^{i \delta} & -s_{23}c_{13}\\
s_{23} s_{12} +c_{12}  s_{13} c_{23} e^{i\delta}&c_{12} s_{23} -  s_{12} s_{13}
c_{23} e^{i \delta} & c_{23}c_{13}\\
\ea \right)\left(
\ba{ccc}
1& & \\
 &e^{i \alpha_2/2}  &\\
 &  &e^{i \alpha_3/2}\\ \ea \right)~.
\ee
Here $c_{ij}=\cos\theta_{ij}, s_{ij}=\sin\theta_{ij}$; $\theta_{ij}$ are three mixing angles,
$\delta$ is Dirac CP phase and $\alpha_{2,3}$ are Majorana CP phases. $m_{1,2,3}$ are three real and
positive neutrino masses. Neutrino (mass)$^2$ differences $\Delta_\odot\equiv m_2^2-m_1^2$ and
$\Delta_A\equiv m_3^2-m_1^2$ and three mixing angles are now experimentally known. The latest global
fit \cite{GonzalezGarcia:2012sz} of neutrino oscillation data gives
\beqa \label{global}
\sin^2\theta_{12}= 0.30\pm 0.013~(0.27-0.34)& ~\Rightarrow~ &
\theta_{12}=33.3\degree \pm 0.8\degree~(31\degree-36\degree) \nonumber \\
\sin^2\theta_{23}=0.41^{+0.037}_{-0.025}\oplus0.59^{+0.021}_{-0.022} ~
(0.34-0.67)& ~\Rightarrow~ &
\theta_{23}={40\degree}^{+2.1\degree}_{-1.5\degree}\oplus{50.4\degree}^{+1.2\degree}_{-1.3\degree}~
(36\degree-55\degree)\nonumber \\
\sin^2\theta_{13}=0.023\pm0.0023~(0.016-0.030)& ~\Rightarrow~ & 
\theta_{13}={8.6\degree}^{+0.44\degree}_{-0.46\degree}~(7.2\degree-9.5\degree)~\nonumber \\
\frac{\Delta_{\odot}}{10^{-5}{\rm eV}^2}=7.5\pm 0.185~(7.0-8.09)& {\rm
and} &\frac{\Delta_A}{10^{-3}{\rm eV}^2}
= 2.47^{+0.069}_{-0.067}~(2.27-2.69),\nonumber \\ \eeqa
where (...) denote the 3$\sigma$ ranges of respective observables. The fit obtains two minima for
$\theta_{23}$ and we choose the one corresponding to $\theta_{23}=40\degree$ for the discussions
presented in this paper.

It follows from Eqs. (\ref{defe1e2}, \ref{mnufgeneral}, \ref{upmns}) that 
\beqa \label{e1e2}
\epsilon_1&=& \frac{y+s_{13} f}{1-s_{13}yf}~,\nonumber \\
\epsilon_2&=& \frac{1}{g_+}\left( (c_{23}^2-s_{23}^2) g_- +4 c_{12}s_{12}c_{23}s_{23}
s_{13}e^{-i \delta}(-m_1+m_2 e^{-i\alpha_2})\right) 
~,\eeqa
with 
\beqa
\label{fg}
f&\equiv&\frac{m_3 e^{-i(\alpha_3-\delta)}-m_1c_{12}^2 e^{-i\delta}-m_2
s_{12}^2e^{-i(\alpha_2+\delta)}}{s_{12}c_{12}(m_1-m_2
e^{-i\alpha_2})}~,\nonumber \\
g_{\pm}&\equiv& \pm m_3 e^{-i\alpha_3}c_{13}^2+m_1 (s_{12}^2\pm  c_{12}^2
s_{13}^2 e^{-2 i \delta})+m_2 e^{-i\alpha_2}(c_{12}^2\pm  s_{12}^2 s_{13}^2
e^{-2 i \delta}) ~.\eeqa
In  the above equations, $s_{13}$
and $y\equiv(c_{23}-s_{23})/(c_{23}+s_{23})$ or equivalently $\frac{1}{2}(c_{23}^2-s_{23}^2)\approx y$ are the \mt
breaking observables which are small and similar in magnitude, see Eq. (\ref{global}): $-0.18\leq
y\leq 0.16$ and $0.12\leq s_{13}\leq 0.17$. Their smallness cannot however be taken as evidence
of an underlying approximate \mt symmetry due to the presence of functions $f$ and $g_\pm$ which
strongly depend on neutrino mass hierarchy and CP violating phases. They can make $\epsilon_1$
and/or $\epsilon_2$ large leading to relatively large breaking of the \mt symmetry. It turns out
that $\epsilon_1$ plays a major role in allowing or disallowing \mt symmetry and we shall
concentrate on it. One could neglect second term in the denominator of $\epsilon_1$ in
Eq. (\ref{e1e2}) for $|f|\ll 75$. In this case, $$\epsilon_1\approx y+s_{13} f~.$$
Thus one can have small $\epsilon_1$ for $|f|\sim {\cal O} (1)$. Let us estimate $f$ for different
neutrino mass hierarchies:\\

\noindent {\bf  (A) Normal hierarchy: $ m_1\ll m_2\approx \sqrt{\Delta_\odot}\ll
m_3 \approx \sqrt{\Delta_A}$}\\
In this case,
\be \label{fnormal}
f\approx - \frac{\sqrt{\Delta_{A}/\Delta_\odot}}{s_{12}c_{12}}
e^{i(\alpha_2-\alpha_3+\delta)}\left( 1+{\cal O}\left(
\frac{\Delta_{\odot}}{\Delta_{A}}\right)\right) ~\Rightarrow~ |f|\approx 12.5~
(1+{\cal O}(0.2))~,\ee
Such an $f$ leads to a large $|\epsilon_1|\approx |y+s_{13} f|\geq 1.5$. Thus, \mnuf~ cannot be
considered to posses an effective \mt symmetry if neutrino mass hierarchy is normal.\\

\noindent {\bf  (B) Inverted  hierarchy:  $ m_1\approx  \sqrt{\Delta_A},~
m_2\approx \sqrt{\Delta_\odot+\Delta_A}\gg m_3 $}\\ 
The values of $\epsilon_{1}$ depend on the CP phases in this case. $f$ can be approximated as
\be 
\label{finverted}
f\approx-e^{-i\delta}\frac{c_{12}^2 + s_{12}^2 e^{-i
\alpha_2}+{\cal O}(\Delta_\odot/\Delta_A)}{s_{12}c_{12}(1-e^{- i
\alpha_2}+{\cal O}(\Delta_\odot/\Delta_A))}~. 
\ee
$|f|$ gets enhanced for $\alpha_2\sim 0$ which results in large $\epsilon_1$ while it is ${\cal
O}(\cot2\theta_{12})$ for $\alpha_2\sim\pi$. Allowed range of $\alpha_2$ is close to
$\pi/2<\alpha_2<\pi$ for which $|\epsilon_1|\leq 0.2$. \\

\begin{figure}[ht!]
\begin{center}
\hspace{-0.8cm}
\includegraphics[width=5.25cm]{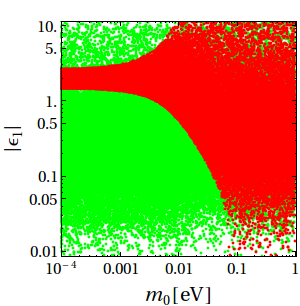}
\hspace{-0.2cm}
\includegraphics[width=5.25cm]{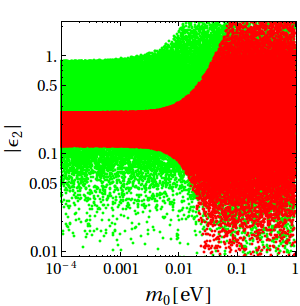}
\hspace{-0.1cm}
\includegraphics[width=5.1cm]{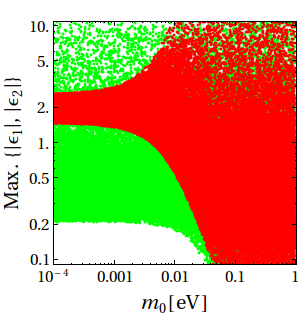}
\caption{Allowed values of $|\epsilon_1|$ (left), $|\epsilon_2|$ (center) and maximum of
$\{|\epsilon_1|,~|\epsilon_2|\}$ (right) as a function of the lightest neutrino mass $m_0$ in case
of the normal (red/dark grey points) and inverted (green/light grey points) hierarchy in neutrino
masses. The scattered points are obtained by varying $\delta,~\alpha_{2,3} \in [0,~2\pi]$ and for
the central values of the other observables as given in Eq. (\ref{global}).}
\label{fig1}\end{center}
\end{figure}

\noindent{\bf  (C) Quasi degeneracy: $m_1=m_0\gg \sqrt{{\Delta_\odot}},~
m_2=\sqrt{m_0^2+\Delta_\odot},~m_3=\sqrt{m_0^2+\Delta_{A}}$}\\ 
An idea of allowed values of $|f|$ can be obtained in this case by considering limiting cases of
the Majorana phases corresponding to CP conserving situations. There are four independent
possibilities with  initial signs of the three masses: (i) + + +, (ii) + + -, 
(iii) + - + and (iv) + - -. The function $f$ in these cases is given by
\beqa
\label{cases}
f&\approx& \frac{\pm (1+\frac{\Delta_A}{2 m_0^2}) e^{i \delta}- e^{-i \delta}
}{-\frac{\Delta_\odot}{2 m_0^2}c_{12}s_{12}} ~~~~~{\rm for~(i)~and~(ii)} ~,
\nonumber \\ 
f&\approx& \frac{1}{\sin2\theta_{12}} \left( \pm \left(1+\frac{\Delta_A}{2
m_0^2}\right)e^{i\delta}-\cos2\theta_{12}e^{-i\delta}\right) ~~~~{\rm
for~(iii)~and~(iv)} ~,
\eeqa
where positive sign refers to cases (i) and (iii) while negative sign refers to cases (ii) and (iv).
It is clear that $|f|$ is very large $\geq \Delta_A/\Delta_\odot$ in cases (i) and (ii) while for
(iii)
and (iv), $|f|<\cot\theta_{12}$ and maximum value is attained for $\delta=\pi/2~(0)$ in case iii
(iv). Both these cases thus allow small $\epsilon_1$. In order to find the range of viability of the
\mt symmetry, one also needs to consider $\epsilon_2$ and allow non-trivial phases. We show the
numerical results of doing this in Fig. \ref{fig1} which displays the values of $|\epsilon_1|$ and
$|\epsilon_2|$ as a function of the lightest neutrino mass in case of the normal and inverted
hierarchy. We also plot in Fig. \ref{fig1} the largest contribution to \mt breaking, namely
Max.$\{|\epsilon_1|,~|\epsilon_2|\}$, for a given mass of the lightest neutrino. 

\begin{figure}[ht!]
\begin{center}
\includegraphics[width=10.5cm]{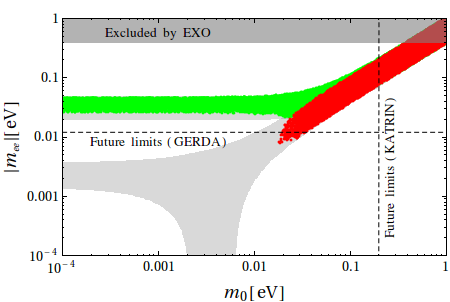}
\caption{Allowed ranges of the effective mass $|m_{ee}|$ as a function of the lightest neutrino mass
$m_0$ in case of the normal (red/dark grey) and inverted (green/light grey) neutrino mass spectrum.
The region covered by the scattered points corresponds to $|\epsilon_{1}|,~|\epsilon_{2}| \le 0.3$
while the shaded region corresponds to the most general case without any restriction on
$|\epsilon_{1,2}|$. The upper grey band shows the region excluded by EXO experiment
\cite{Auger:2012ar} and the horizontal and vertical dashed lines correspond to the future
sensitivity of relevant experiments.}
\label{fig2}
\end{center}
\end{figure}

It is seen from Fig. \ref{fig1} that the largest contribution to \mt breaking comes
from $|\epsilon_{1}|$ in case of the normal hierarchy. A small violation of \mt symmetry, less than
20\%, disfavors hierarchical neutrino spectrum ($m_0<0.025$ eV) in this case.
The inverted hierarchy allows small values of $|\epsilon_{1}|$ or $|\epsilon_{2}|$ but both of them
are not simultaneously small. In this case, one is able to have small \mt breaking, {\it i.e.}
Max.$(|\epsilon_1|,~|\epsilon_2|)\leq {\mathcal O}(0.2)$ even for $m_3$ close to zero. Thus, only
quasidegenerate or inverted neutrino spectrum provides a viable alternative for the \mt
symmetry to remain an approximate symmetry of \mnuf.\footnote{Similar conclusion has been reached
earlier \cite{probir} in a specific context of mass matrices with texture zeros.} This has direct
implications in terms of observables namely, the effective mass $m_{ee}$, the electron neutrino mass
$m_e$ and sum of the neutrino masses as would be inferred from direct mass determination and
cosmology.
A small violation of $\mu$-$\tau$ symmetry corresponding to $|\epsilon_{1,2}| \le 0.3$ leads to the
following predictions for these observables:
\beqa \label{numass_limits}
\lvert m_{ee} \rvert & \equiv & \lvert \sum U_{ei}^2 m_i \rvert \ge 0.01~\text{eV}, \nonumber \\
m_{e} & \equiv & \sqrt{\sum \lvert U_{ei}\rvert^2 m_i^2} \ge 0.02~\text{eV}, \nonumber \\
m_{\rm cosmo.} & \equiv & \sum m_i \ge 0.1~\text{eV}. \eeqa
Of these, we show the allowed region of $|m_{ee}|$ as a function of the lightest
neutrino mass in Fig. \ref{fig2}.  
As it can be seen, the region in $|m_{ee}|$ corresponding to the normal
hierarchy is strongly disfavored if \mt symmetry is to remain viable in a way discussed here.
 In particular, non-observation of signal in experiments like GERDA \cite{Smolnikov:2008fu}, CUORE
\cite{Giachero:2012zz} and MAJORANA \cite{MAJORANA:2011aa} (also see \cite{Rodejohann:2012xd} for the recent
review) would practically rule out approximate \mt symmetry only
of \mnuf~as a possible explanation behind the small value of $\theta_{13}$.

\section{Approximately $\mu$-$\tau$ symmetric $M_\nu$ and $M_l$}
\begin{figure}[ht!]
\begin{center}
\hspace{-0.8cm}
\includegraphics[width=5.3cm]{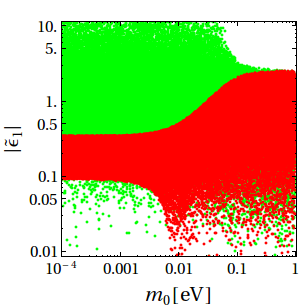}
\hspace{-0.2cm}
\includegraphics[width=5.15cm]{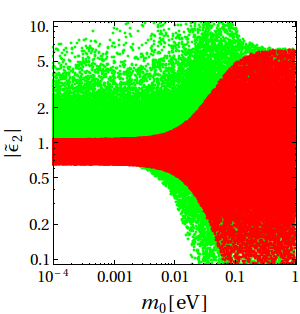}
\hspace{-0.1cm}
\includegraphics[width=5.15cm]{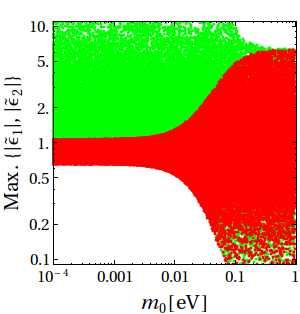}
\caption{Allowed values of $|\tilde{\epsilon}_1|$ (left) and $|\tilde{\epsilon}_2|$ (center) and
maximum of $\{|\tilde{\epsilon}_1|,~|\tilde{\epsilon}_2|\}$ (right) as a function of the lightest
neutrino mass $m_0$ in case of the normal (red/dark grey points) and inverted (green/light grey
points) hierarchy in neutrino masses. The scattered points are obtained by varying
$\delta,~\alpha_{2,3} \in [0,~2\pi]$ and for the central values of the other observables as given in
Eq. (\ref{global}).}
\label{fig3}\end{center}
\end{figure}

\mt symmetry of \mnuf~need not imply it's presence at the fundamental level. A well-known example
is $A_4$ group imposed as a symmetry of the Lagrangian. This does not even contain \mt symmetry as a
subgroup but its spontaneous breaking in a specific manner leads to an \mnuf~displaying \mt
symmetry \cite{reviews}. One could take an alternative point of view and regard \mt symmetry itself as more
fundamental. We shall now explore phenomenological viability of this scenario. To this end we
start by assuming that both the charged lepton mass matrix $M_l$ and $M_\nu$ are simultaneously \mt
symmetric in a suitable basis. More specifically, we assume,
\beqa\label{mlmnu}
S_2^TM_\nu S_2&=&M_\nu ~,\nonumber \\
S_2^\dagger M_l M_l^\dagger S_2&=& M_l M_l^\dagger ~, \eeqa
with $S_2$ defined in Eq. (\ref{s2}). The $M_lM_l^\dagger $ is diagonalized by
\be \label{mld}
U_l^\dagger M_lM_l^\dagger U_l=D_l ~\ee
$D_l$ is a  diagonal matrix and  
\be \label{r}
U_{l}=R_{23}(\pi/4) U_{12}~. \ee
where $R_{23}(\pi/4)$ denotes ordinary rotation in the 23 plane by an angle $\pi/4$ while $U_{12}$
is a general unitary rotation in the 12 plane. It follows from Eqs. (\ref{mlmnu}, \ref{mld}) that
${\cal M}_{\nu f}\equiv U_l^T M_\nu U_l$ satisfies
\be \label{s2tilde}
\tilde{S}_2^T {\cal M}_{\nu f} \tilde{S}_2={\cal M}_{\nu f} ~,\ee
where 
$$ \tilde{S}_2\equiv U_l^\dagger S_2 U_l={\rm Diag.}(1,1,-1).$$
showing that imposition of the \mt symmetry on $M_l$ and $M_\nu$ is equivalent to imposing
$\tilde{S_2}$ on \mnuf. Thus one should demand \mnuf~to be invariant under $\tilde{S}_2$ {\it and
not} \mt symmetry if the latter is to arise at the fundamental level. The most general
\mnuf~invariant under $\tilde{S_2}$ can be written as
\be \label{mnuftilde0}
{\cal M}_{\nu f}^0=\left(
\ba {ccc}
x&a&0\\
a&b&0\\
0&0&c\\
\ea \right)~.\ee
This form and hence the exact $\tilde{S}_2$ invariance is clearly not a viable proposition since
it allows only the solar mixing angle to be non-zero. One must therefore break it. Admitting
symmetry breaking, \mnuf~can be written as
\be \label{mnuftilde}
{\cal M}_{\nu f}=\left(
\ba {ccc}
x&a&\tilde{\epsilon}_1 c\\
a&b&\tilde{\epsilon}_2 c\\
\tilde{\epsilon}_1c&\tilde{\epsilon}_2c&c\\
\ea \right)~,\ee
where $\tilde{\epsilon}_{1,2}$ parameterize the symmetry breaking. We have normalized them with
respect to the (3,3) element of ${\cal M}_{\nu f}$\footnote{In Eq. (\ref{mnuftilde}), $c$ turns out to be the
largest element in case of the normal hierarchy. In case of inverted ordering, it is almost degenerate with the largest
element in ${\cal M}_{\nu f}$ as can be seen from Eqs. (\ref{evs}). Also in case of the quasidegenerate
spectrum, $c$ is nearly equal to or is the largest element in ${\cal M}_{\nu f}$.}. We now try to find out under
what circumstances
$\tilde{\epsilon}_{1,2}$ can be small. As before, we  express these parameters in terms of
observables by comparing Eq. (\ref{mnufgeneral}) with the form of \mnuf~given in Eq.
(\ref{mnuftilde}). This leads to
\beqa 
\tilde{\epsilon}_1&=&\frac{c_{13}c_{12}s_{12} (m_1-m_2 e^{-i \alpha_2})\left(-s_{13}c_{23} f
+s_{23}\right)}{c_{23}^2g_+ -\cos2\theta_{23}(m_2 e^{-i\alpha_2}
c_{12}^2+m_1 s_{12}^2)+\sin2 \theta_{23} c_{12} s_{12} s_{13} e^{-i
\delta}(m_1-m_2 e^{-i\alpha_2})}~,\nonumber \\
\tilde{\epsilon}_2&=&\frac{c_{23}s_{23} g_- +\cos2\theta_{23} c_{12}s_{12}
s_{13}e^{-i\delta}(m_1-m_2 e^{-i \alpha_2})}{c_{23}^2g_+ -\cos2\theta_{23}(m_2
e^{-i\alpha_2} c_{12}^2+m_1 s_{12}^2)+\sin2 \theta_{23} c_{12} s_{12} s_{13}
e^{-i \delta}(m_1-m_2 e^{-i\alpha_2})}~.\nonumber \\
\eeqa
where $f$ and $g_\pm$ are defined in Eq. (\ref{fg}). The magnitudes of these parameters are plotted
as a function of the lightest neutrino mass in Fig. \ref{fig3} for normal and inverted hierarchy.
As can be seen form Fig. \ref{fig3},
\begin{itemize}
 \item  For both the normal and inverted hierarchies, $|\tilde{\epsilon}_1|$ and
$|\tilde{\epsilon}_2|$ remain small ($< 0.2$) only if $m_0>0.04$ eV. Thus one cannot regard
$\tilde{S}_2$ as an approximate symmetry of \mnuf~in these two cases.
 \item In contrast, for the quasidegenerate spectrum, $\tilde{S}_2$ and hence $S_2$ at the
fundamental level can be an approximately good symmetry. This was argued earlier in
\cite{Joshipura:2005vy} and it can be seen analytically as follows. The diagonalization of Eq.
(\ref{mnuftilde}) yields in the approximation of neglecting terms of ${\cal O}(s_{13}^2,a s_{13})$ and assuming real 
parameters,
\beqa
\label{evs}
\tan 2\theta_{23}&\approx &\frac{2 c \tilde{\epsilon}_2}{b-c} ~,\nonumber \\
\tan 2\theta_{12}&\approx &\frac{2(a c_{23}+ c\tilde{\epsilon}_1
s_{23})}{m_2-x} ~,\nonumber \\
\tan 2\theta_{13}&\approx &\frac{2( c\tilde{\epsilon}_1 c_{23}-a s_{23})}{m_3-x}
~,\nonumber \\
m_3&\approx &\frac{1}{2}\left (b+c-\frac{b-c}{\cos2
\theta_{23}}\right),\nonumber\\
m_2&\approx &\frac{1}{2} \left(b+c+\frac{b-c}{\cos2
\theta_{23}}\right),\nonumber\\
m_1&\approx &\frac{1}{2} \left(x+m_2+\frac{x-m_2}{\cos2
\theta_{12}}\right).\eeqa
As seen from above, a  large atmospheric mixing is consistent with a small $\tilde{\epsilon}_2$ for
$b\approx c\gg \tilde{\epsilon}_2$ which corresponds to $m_2\sim b+c\tilde{\epsilon}_2,~m_3\sim
b-c\tilde{\epsilon}_2$. $m_1$ is then required to be degenerate if both solar and atmospheric
neutrino scales are to be reproduced. In contrast, for $c\ll b$ or $b\ll c$, $\tilde{\epsilon}_2$ is
forced to be ${\cal O}(1)$ and one needs a large \mt breaking.
\end{itemize}

Given these restrictions, it is indeed possible to choose parameters in Eq. (\ref{mnuftilde}) which
reproduce all the observables correctly. To show this, we try to fit parameters in Eq.
(\ref{mnuftilde}) in two different ways. In the first, we minimize relevant $\chi^2$ by restricting
$|\tilde{\epsilon}_{1,2}|$ to be $\leq 0.1$. This leads to the following solution
\be
{\cal  M}_{\nu f}=0.07683~{\rm eV}
\left(
\begin{array}{ccc}
 0.88253 & -0.01772 & 0.02639 \\
 -0.01772 & 0.96430 & -0.1 \\
 0.02639 & -0.1 & 1 \\
\end{array}
\right)~, \ee
which corresponds to $\chi^2\approx10^{-2}$ at the minimum and reproduces the central
values of $\theta_{23},~\theta_{13},~\theta_{12}$ and $\Delta_\odot/\Delta_A$. The overall mass  is
normalized to get the correct atmospheric scale. This leads to the following neutrino masses:
\be
(m_1,~m_2,~m_3)=(0.06726,~0.06782,~0.08363)~{\rm eV}.~\ee
Restricting $\tilde{\epsilon}_{1,2}$ to small values automatically leads to quasidegenerate
spectrum as would be expected. In contrast, performing  the same fit without putting any
restrictions on $\tilde{\epsilon}_{1,2}$ led to 
\be
{\cal  M}_{\nu f}=0.03189~{\rm eV}
\left(
\begin{array}{ccc}
 \text{0.03992} &
   \text{0.2954} &
   \text{0.05802} \\
 \text{0.2954} &
   \text{0.6988} &
   \text{0.6615} \\
 \text{0.05802} &
   \text{0.6615} &
   \text{1}
\end{array}
\right)~,\ee
corresponding to the minimum $\chi^2\approx10^{-3}$. This gives  correct central values of all
observables and neutrino masses
\be 
(m_1,~m_2,~m_3)=(0.00388,~0.00948,~0.04985)~{\rm eV}.\ee
corresponding  to a normal spectrum. This however requires large symmetry breaking
$\tilde{\epsilon}_2\sim 0.66$ as would be expected.

\section{Approximate \mt symmetric Lagrangian}
\label{approximate}
So far we have assumed  mass matrices ${\cal M}_{\nu f}$  alone or $M_l$ and $M_\nu$ to be approximately \mt symmetric.
We now discuss the circumstances under which this  symmetry may originate from the symmetry in the underlying  Lagrangian.
We motivate it through a simple example \cite{Joshipura:2005vy,Joshipura:2007sf}  containing  two Higgs doublets 
$\phi_{1,2}$. $\phi_1$ $(\phi_2)$ is assumed even (odd) under the \mt symmetry. The Yukawa couplings of the charged
leptons then have the following form:
\be\label{yukawa}
 -{\cal L}_Y=\overline{l_L} (\Gamma_1\phi_1+\Gamma_2\phi_2)e_R+{\rm h.c.}~. \ee
 $l_L,e_R$ respectively denote three generations of the leptonic doublets and singlets. Yukawa
couplings $\Gamma_{1,2}$ satisfy $S_2^T\Gamma_1S_2=\Gamma_1$ and $S_2^T \Gamma_2 S_2=-\Gamma_2$.
Approximately \mt symmetric $M_l$ would result from the above if
$|\frac{(\Gamma_2)_{ij}}{(\Gamma_1)_{ij}}||\frac{<\phi_1^0>}{<\phi_2^0>}|\ll 1$. Situations with
$M_\nu$ is however different. If neutrino masses result from the type-II seesaw mechanism with
direct coupling of one or more triplet Higgs to neutrinos then just like $M_l$, $M_\nu$ would also
display an approximate \mt symmetry. In this case, as shown above \mt symmetry can be approximate
only for the quasidegenerate spectrum and hierarchical mass spectrum is inconsistent
with it. In contrast, if neutrinos obtain their masses from the type-I seesaw mechanism then the
Dirac mass matrix $m_D$ would originate from the Yukawa couplings similar to Eq. (\ref{yukawa}) and
will display an approximate \mt symmetry. The explicit Majorana mass matrix $M_R$ for the right
handed neutrinos  appears directly in the Lagrangian and would be \mt symmetric when this symmetry
is imposed on the Lagrangian. The resulting neutrino mass matrix $M_\nu\approx -m_D M_R^{-1} m_D^T$
may however contain large breaking of the \mt symmetry even when $m_D$ and $M_R$ are approximately
\mt symmetric. Thus in type-I seesaw mechanism the normal or inverted hierarchical neutrino spectrum
can also be consistent with the approximately \mt symmetric Lagrangian. This was realized and
discussed in detail in \cite{Joshipura:2005vy}. Here let us illustrate it with a simple but
sufficiently realistic example.

\begin{figure}[ht!]
\begin{center}
\includegraphics[width=10.5cm]{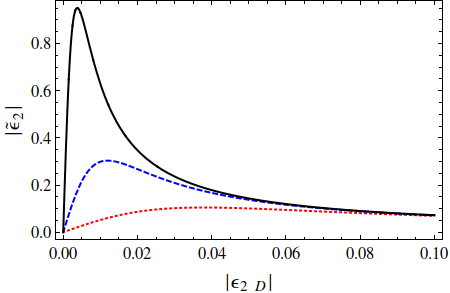}
\caption{The parameter $\tilde{\epsilon}_{2}$ as a function of $\epsilon_{2D}$ for $\frac{M_2}{M_3} = 0.1
\frac{m_{2D}}{m_{3D}}$ (dotted red), $\frac{M_2}{M_3} = 0.01 \frac{m_{2D}}{m_{3D}}$ (dashed blue) and $\frac{M_2}{M_3} =
0.001 \frac{m_{2D}}{m_{3D}}$ (solid black).}
\label{fig4}
\end{center}
\end{figure}

Let us assume that $M_R$ and $M_l$ are \mt symmetric and small breaking of this symmetry occurs only
in the Dirac neutrino mass matrix $m_D$ which is assumed symmetric. The latter is thus parameterized by
\be \label{mdirac}
m_D=\left(
\ba{ccc}
x_D&a_D(1-\epsilon_{1D})&a_D(1+\epsilon_{1D})\\
a_D(1-\epsilon_{1D})&b_D(1-\epsilon_{2D})&c_D\\
a_D(1+\epsilon_{1D})&c_D&b_D(1+\epsilon_{2D})\\
\ea \right)~,\ee
Here, $\epsilon_{1D,2D}$ are small \mt symmetry breaking parameters. As discussed in
\cite{Joshipura:2005vy}, if the eigenvalues of $m_D$ and $M_R$ are hierarchical and if hierarchy in
the right handed neutrino masses are stronger such that the $M_R$ is nearly singular
\cite{Smirnov:1993af} then the resulting $M_\nu$ may show large breaking of \mt symmetry. $M_l$ and
$M_R$ are diagonalized by a matrix of the form $R_{l,R}=R_{23}(\pi/4) R_{12}(\theta_{12l,R})$.
Assuming small $\theta_{12l,R}$, neutrino mass matrix in the flavour basis can be written as
\be \label{mnuftype1}
{\cal M}_{\nu f}\approx  \tilde{m}_D^T ~{\rm Diag.} (M_1^{-1},M_2^{-1},
M_3^{-1})~\tilde{m}_D ~,\ee
with $\tilde{m}_D\equiv R_{23}^T(\pi/4)~ m_D~R_{23}(\pi/4) $.  Comparing this
with Eq. (\ref{mnuftilde}), one finds to leading order in $\epsilon_{1D,2D}$,
\beqa \label{e1e2tilde}
\tilde{\epsilon}_1&\approx & \frac{\sqrt{2} a_D  (\epsilon_{2D} b_D M_1
   M_3+\epsilon_{1D} M_2 (b_D M_1-c_D
   M_1+M_3 x_D))}{(b_D-c_D)^2 M_1
   M_2}~,\nonumber\\
   \tilde{\epsilon}_2&\approx & \frac{ 2 \epsilon_{1D} M_2 M_3
   a^2_D+\epsilon_{2D} b_D M_1 (c_D
   (M_3-M_2)+b_D
   (M_2+M_3))}{(b_D-c_D)^2 M_1
   M_2}~.\eeqa
  As shown above, neutrino mass hierarchy $m_1\ll m_2\ll m_3$ requires small $\tilde{\epsilon}_1$
and relatively large $\tilde{\epsilon}_2$ (see, Fig. \ref{fig3}). This can be reconciled with a
small breaking, {\it i.e.} $|\epsilon_{1D,2D}|\ll1 $ at the fundamental level. Let us assume
hierarchical eigenvalues $m_{1D}\ll m_{2D}\ll m_{3D}$ for the Dirac mass matrix $m_D$. This can
result with $x_D\ll a_D\sim \sqrt{m_{1D}m_{2D}}\ll b_D,c_D$. In this case, to the leading order in 
$\epsilon_{1D,2D}$ one has $b_D\approx \frac{1}{2}(m_{2D}+m_{3D})$ and $c_D\approx 
\frac{1}{2}(m_{2D}-m_{3D})$. Inserting these in Eq. (\ref{e1e2tilde}), one gets
\be \label{ele2tilde_simplified}
\frac{\tilde{\epsilon}_1}{\tilde{\epsilon}_2}\approx
\frac{\sqrt{2m_{1D}/m_{2D}}}{1+\frac{M_2}{M_3} \frac{m_{2D}}{m_{3D}}}~ ~~~{\rm and}~~~
\tilde{\epsilon}_{2} \approx \frac{\epsilon_{2D}}{2} \left(1+\frac{m_{2D}}{m_{3D}} \right)
\left(1+\frac{m_{2D} M_3}{m_{3D} M_2} \right)~. \ee
We have assumed $\epsilon_{1D}\ll \epsilon_{2D}$ and neglected contribution of $\epsilon_{1D}$ in
writing the above equation. Strong RH mass hierarchy $\frac{M_2}{M_3}\ll \frac{m_{2D}}{m_{3D}}$ and
hierarchical $m_{iD}$ automatically lead to enhancement in $\tilde{\epsilon}_{2}$ compared to the
basic parameter $\epsilon_{2D}$ and the ratio $\tilde{\epsilon}_{1}/\tilde{\epsilon}_2$
remains small as required. 
For illustration, we take $m_{2D}/m_{3D} \approx m_c/m_t\approx 3.6\times
10^{-3}$, $b_D=\frac{1}{2}(m_{2D}+m_{3D})$, $c_D=\frac{1}{2}(m_{2D}-m_{3D})$ and $a_D,~x_D,~\epsilon_{1D} \approx 0$ in
Eq. (\ref{mdirac}) and evaluate $\tilde{\epsilon}_{2}$ using Eq. (\ref{mnuftype1}) for the different values of
$M_2/M_3$. The results are displayed in Fig. \ref{fig4}.
As can be seen, one obtains large enough $\tilde{\epsilon}_{2}$ for small values of $\epsilon_{2D}$ in case of large
hierarchy in RH neutrino masses. $\tilde{\epsilon}_{1}$ is suppressed by factor $\sqrt{2m_{1D}/m_{2D}}$ and remains
small as desired.

\section{Summary}
We have systematically investigated impact of the measurement of the reactor mixing angle
$\theta_{13}$ on the viability of the \mt symmetry. The first investigated scenario is the
standard one \cite{mt}  in which \mt symmetry is imposed as an effective symmetry of \mnuf~only and
the
charged lepton mass matrix does not respect it. Admitting general symmetry breaking, we found that
the symmetry breaking parameters can be small only if the neutrino spectrum is inverted or
quasidegenerate. This leads to direct prediction that the neutrinoless double beta decay should be
in the observable range. In the second scenario, we assumed both $M_l$ and neutrino mass matrix
$M_\nu$ to be \mt symmetric. This is equivalent to imposing the $\tilde{S}_2$ symmetry Eq.
(\ref{s2tilde}), on \mnuf. The diagonal charged lepton mass matrix also remains invariant under
this. Again, admitting symmetry breaking, one reaches conclusion that the \mt symmetry imposed on
$M_l,~M_\nu$ is viable as an approximate symmetry only for the quasidegenerate spectrum.

In either scenario, the hierarchical neutrino masses imply large breaking of \mt symmetry. If
neutrinos obtain their masses from the type-II seesaw mechanism then such large breaking would not
allow \mt symmetry to be interpreted as a symmetry of the underlying Lagrangian. In contrast, the
type-I seesaw mechanism allows interesting possibility in which the required large breaking may be
understood as a seesaw amplification of small symmetry breaking in the underlying  Lagrangian. This
is illustrated in Section \ref{approximate} and is discussed at length in \cite{Joshipura:2005vy}.

To sum up, \mt symmetry in either of the presented scenarios is viable as an approximate symmetry in
type-II seesaw only if neutrino spectrum is inverted or quasidegenerate in  nature. Type-I
seesaw mechanism allows also the normal hierarchy in neutrino masses and a small breaking
at the fundamental level.\\

\acknowledgments
ASJ thanks the Department of Science and Technology, Government of India for support under the J. C.
Bose National Fellowship programme, grant no. SR/S2/JCB-31/2010.

\end{document}